\documentclass[aps,prl,twocolumn,showpacs,superscriptaddress,letterpaper,asmmath,amssymb]{revtex4}  

\usepackage{graphicx}
\usepackage{dcolumn}
\usepackage{bm}

\RequirePackage{xspace}
\usepackage{relsize}

\def\Kbar  {\kern 0.2em\overline{\kern -0.2em K}{}\xspace}

\def\Bbar    {\kern 0.18em\overline{\kern -0.18em B}{}\xspace}

\def\Qbar    {\kern 0.08em\overline{\kern -0.08em Q}{}\xspace}

\newcommand{\mev}{\ensuremath{\mathrm{\,Me\kern -0.1em V}}\xspace}
\newcommand{\mevc}{\ensuremath{{\mathrm{\,Me\kern -0.1em V\!/}c}}\xspace}
\newcommand{\mevcc}{\ensuremath{{\mathrm{\,Me\kern -0.1em V\!/}c^2}}\xspace}
\newcommand{\gev}{\ensuremath{\mathrm{\,Ge\kern -0.1em V}}\xspace}
\newcommand{\gevc}{\ensuremath{{\mathrm{\,Ge\kern -0.1em V\!/}c}}\xspace}
\newcommand{\gevcnospace}{\ensuremath{{\mathrm{\,Ge\kern -0.1em V\!/}c}}}
\newcommand{\gevcc}{\ensuremath{{\mathrm{\,Ge\kern -0.1em V\!/}c^2}}\xspace}

\begin{document}
\bibliographystyle{apsrev}

\title{Experimental Sensitivity for Majorana Neutrinos \\Produced via a Z Boson at Hadron
  Colliders}

\author{A.~Rajaraman}
\author{D.~Whiteson}
\affiliation{University of California, Irvine, Irvine, California 92697}

\begin{abstract}
We present experimental sensitivity for the production of a
fourth-generation Majorana neutrino ($N$) via an $s$-channel $Z$
boson $Z\rightarrow NN \rightarrow W\ell W\ell$ at hadron colliders. This channel
has not been studied for the Tevatron  dataset with $\int
\mathcal{L}=5 $\ fb$^{-1}$, where we find
that a single experiment could significantly extend current 95\%
C.L. mass lower limits, to
$m_N>175$ GeV$/c^2$, or report $3\sigma$ evidence for the $N$ if $m_N<150$
GeV$/c^2$. With 5 fb$^{-1}$, a single LHC experiment at $\sqrt{s}=10$ TeV could expect to set a 95\%
C.L. mass lower limits of $m_N>300$ GeV$/c^2$, or report 3$\sigma$
evidence for the $N$ if $m_N<225$ GeV$/c^2$.
\end{abstract}



\maketitle

\date{\today}

\section{Introduction}

The study of a fourth generation of quarks and leptons has undergone a renaissance.
While earlier studies \cite{Amsler:2008zzb} claimed that the fourth generation
was ruled out by experimental data, a recent
analysis~\cite{Kribs:2007nz} has shown that
the constraints can be satisfied by appropriate choices of
mass splittings between the new fourth generation
particles. Furthermore, the existence of the fourth generation can help
 to address certain discrepancies between the Standard Model and experimental results
in the b-quark
sector\cite{Aaltonen:2007he,:2008fj,Hou:2005hd,Hou:2005yb,Soni:2008bc,:2008zza},
leading to a revival of interest in this
possibility (see \cite{Holdom:2009rf} for a review).

In view of the imminent arrival of LHC data, it is of particular interest to
search for signals of these fourth-generation fermions at hadron colliders. Such analyses
have been performed for the $t'$ and $b'$ quarks.
CDF has searched for the $t'$ quark~\cite{Cox:2009mx} using the
CKM suppressed decay $t'\rightarrow qW$  and for the
$b'$ quark~\cite{Aaltonen:2009nr} using the process $b'b'\rightarrow t\bar{t}W^-W^+$
which can yield like-sign dileptons. These searches have set
limits of 311 GeV for the $t'$~\cite{Cox:2009mx}, and 338 GeV  for the $b'$~\cite{Aaltonen:2009nr} (but see \cite{Hung:2007ak}). The LHC will discover or exclude fourth generation quarks up to about a TeV \cite{Ozcan:2008zz,Cakir:2008su,Holdom:2007ap}.

However, few detailed analyses have been performed in the lepton sector.
By analogy with the first three generations, one might expect the
leptons of the fourth generation to be lighter than the fourth generation
quarks and in particular, the fourth generation neutrino might be expected to
be the lightest new particle. It is of interest to see if this neutrino
can be found at colliders.

Past searches for fourth generation neutrinos have mainly been performed
at lepton colliders. In particular, LEP II has looked for neutrinos produced
in the process  	$e^+e^-\rightarrow Z\rightarrow NN$, where the
neutrinos subsequently decay via the process $N\rightarrow l^{\pm}W^{\mp}$.
No excess of such events was found, which placed a limit
of about 100 GeV for Dirac neutrinos decaying to electrons, muons or
taus. For Majorana neutrinos the corresponding limits were
about 90 GeV if the neutrino decayed either to electrons
or to muons, and about 80 GeV if it decayed to taus.

Theoretical analyses of fourth generation neutrinos at hadron colliders \cite{Han:2006ip,Atre:2009rg,delAguila:2008ir,del Aguila:2007em}
have focused on the process
$q\bar{q}'\rightarrow W^{\pm}\rightarrow Nl^{\pm}$ where the fourth generation
neutrino is produced in association with a light charged lepton. This
process has the significant advantage that only one heavy particle is produced,
which increases the mass reach considerably. Furthermore, the
neutrino will decay through $N\rightarrow l^\pm W^{\mp}$ which will
produce the low-background like-sign dilepton signature in half the events.

However, the production cross-section for this process depends on the
mixing between the fourth generation with the first three generations
due to the $W\rightarrow lN$ vertex. In many models, this mixing angle
can be small; for example, if the mixing is generated by GUT or
Planck-scale suppressed operators, the angle may be as small as
$10^{-11}$ \cite{Hung:2007ak}. For mixing parameters smaller than
about $10^{-6}$, the neutrino production rates in this channel are too
small to be observable at colliders~\cite{Han:2006ip}.
In models with small mixing angles, the dominant production mechanism becomes pair production
through an $s$-channel $Z$, for which the production rate is model-independent. 
These signals have been studied at various benchmark points for the LHC~\cite{CuhadarDonszelmann:2008jp}
and for future linear colliders~\cite{Ciftci:2005gf}.

However, the analysis of the $s$-channel $Z$ process has not been performed for the Tevatron. 
In this Letter, we present a sensitivity study for the Tevatron and argue
that the LEP limits on Majorana neutrinos can be significantly improved with an analysis of
the data already taken. It would be of great interest to perform a full analysis
of this data. We also perform a similar study for the LHC, which
can probe the parameter space to much higher energies.

\section{Production and Decay}

We consider an extension to the standard model by a fourth generation
of fermions and a right-handed neutrino.  The mass term for the neutrinos can be written as
\begin{eqnarray}
L_m=-{1\over 2}\overline{(Q_R^c
N_R^c)}\left(\begin{array}{cc}0&m_D\\m_d&M\end{array}\right)
\left(\begin{array}{c}Q_R\\ N_R\end{array}\right)+h.c.
\end{eqnarray}
where $\psi^c=-i\gamma^2\psi^*$.
This theory has two mass eigenstates of masses
 $m_1=-(M/2)+ \sqrt{m_D^2+{M^2/4}},
m_2=(M/2)+ \sqrt{m_D^2+{M^2/4}}$. In addition, the mass of the fourth
generation lepton is constrained to be close in mass to the neutrinos
by precision electroweak constraints~\cite{Kribs:2007nz}.

We consider processes where
only the lightest neutrino is produced, providing the most model-independent bound
on this theory.  In future work, we will study the effect of the
second neutrino and fourth generation lepton. For this analysis, we treat the lepton and second
neutrino as infinitely massive, corresponding to a limit where
$M, m_D$ go to infinity with ${m_D^2\over M}$ fixed.

The lighter neutrino mass eigenstate is the Majorana fermion
$N= N_L^c+ N_L$.  The coupling for pair production via the $Z$ is
through the coupling $L_Z=Z_\mu J^\mu$ where

\[ J^\mu={e\over 2\sin\theta_W\cos\theta_W}(\bar N_1 \gamma^\mu \gamma^5N_1)\]

\begin{figure}[h]
\includegraphics[width=0.6\linewidth]{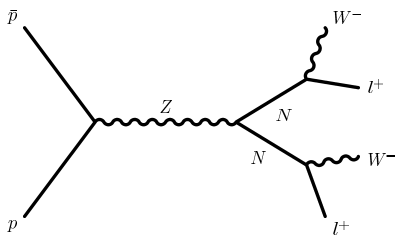}
\caption{ Pair production of the heavy Majorana neutrino $N$ via a $Z$
  boson, and subsequent decay $W^{\pm}l^{\mp}$. }
\label{fig:prod}
\end{figure}

The heavy neutrino will decay  through $N\rightarrow W^{\pm}l^{\mp}$
(the neutral current process is suppressed.) Note that $N$ can decay to
either sign of lepton, giving like-sign leptons in half of the events,
see Fig~\ref{fig:prod}. We assume that the heavy neutrino decays
promptly; this will be the case unless the mixing between the fourth
generation and the first three is extremely small \cite{Hung:2007ak}.

We consider the possible decay modes $N\rightarrow
W(e,\mu,\tau)$. In a hadron
collider, backgrounds to $\tau$ leptons are much larger and efficiencies are
much lower than for $e$ and $\mu$, giving the $\tau$ decay mode little
power. We consider two cases, (a) $\mu\mu$, in which the non-$\tau$ decays appear solely as muons: BR($N\rightarrow W\mu$) = 1 -
  BR($N\rightarrow W\tau$); and (b) $\ell\ell$, with $\ell=e,\mu$ in which the
  non-$\tau$ decays appear as both electrons and
  muons:  BR($N\rightarrow W\mu$) +  BR($N\rightarrow We$) = 1 -
  BR($N\rightarrow W\tau$). The  $\mu^{\pm}\mu^{\pm}$ mode has
  significantly smaller background rate than  $\ell^{\pm}\ell^{\pm}$.

If the $N$ decays to $\ell W$, then the decay of
$NN \rightarrow \ell W\ell W$ can be categorized by the
decays of the $W$ bosons.  If both $W$s decay hadronically, we expect
approximately 4 jets. If one decays leptonically, we expect
approximately 2 jets and
a third lepton. If both decay leptonically, we expect approximately
zero jets but four leptons.  All but the fully leptonic mode, the smallest fraction,
contribute to the $\ell^{\pm}\ell^{\pm}jj$ signature and allow for
direct reconstruction of the $N$.

\section{Experimental Sensitivity}

We select events with the $\ell^{\pm}\ell^{\pm}jj$ signature:
\begin{itemize}
\item two like-signed reconstructed leptons ($e$ or $\mu$), each with $p_T > 20$ GeV
  and $|\eta|<2.0$
\item at least two reconstructed jets, each with $p_T > 15$ GeV
  and $|\eta|<2.5$
\end{itemize}

\begin{figure}[h]
\includegraphics[width=0.8\linewidth]{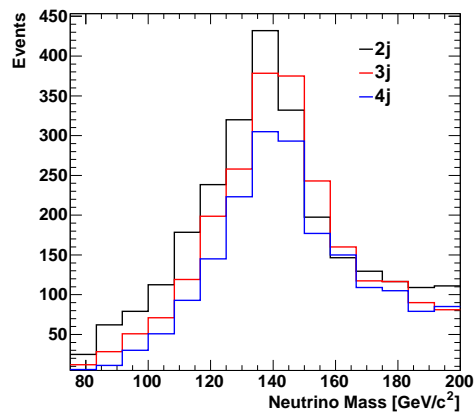}
\caption{Reconstructed Majorana neutrino ($N$) mass in events with 2, 3 or at least 4
jets for $m_N=150$\ GeV/$c^2$.}
  \vspace{-3ex}
\label{fig:masses}
\end{figure}

In the case that two jets are reconstructed, the mass of the $N$ can
be reconstructed from the two jets and either of the leptons. In
the case that three jets are reconstructed, we form the mass of the
$N$ from the invariant mass of the two jets which are closest to the $W$ mass,
and either of the leptons.  In the case that four jets are
reconstructed, the mass of the  two $N$s come from the $ljj$
assignments that give the best $W$ masses and the smallest difference
between the two  reconstructed $m_{N}$s.  See Figure~\ref{fig:masses}.

\subsection{Backgrounds}

At the Tevatron, the largest backgrounds to the  $\ell^{\pm}\ell^{\pm}jj$ signature come
from $W\gamma$ or $WZ$ production or misidentified leptons~\cite{cdf1fbprl} either from
semi-leptonic $t\bar{t}$ decays or direct $W+$ jets production.

For the Tevatron, we extrapolate the number of expected backgrounds
events from Ref.~\cite{cdf1fbprl} to a dataset with 5 fb$^{-1}$, use {\sc madgraph}~\cite{madgraph} to model the kinematics
of the events, {\sc pythia}~\cite{pythia} for showering and a version of
{\sc pgs}~\cite{pgs} tuned to describe the performance of the CDFII detector.

At the LHC, the diboson contribution includes an additional
process, $qq\rightarrow W^{\pm}W^{\pm}q'q'$, which directly produces the $\ell^{\pm}\ell^{\pm}jj$ signature. 
For the LHC, we calculate the size and kinematics of each contribution
using {\sc madgraph}, use {\sc pythia} for showering and a version of {\sc
  pgs} tuned to describe the performance of the ATLAS detector.

\begin{figure}[h]
\includegraphics[width=0.8\linewidth]{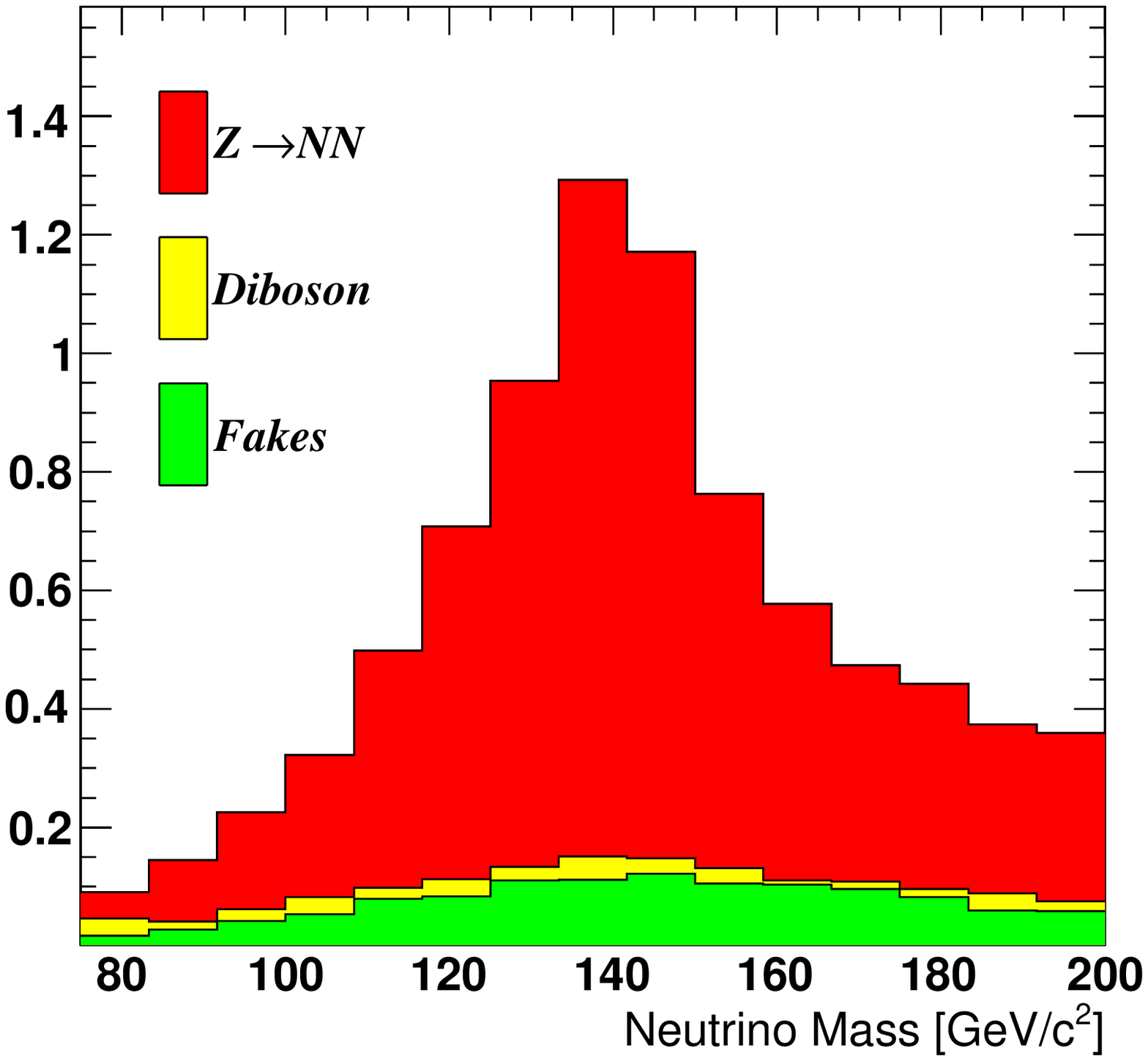}\\
\includegraphics[width=0.8\linewidth]{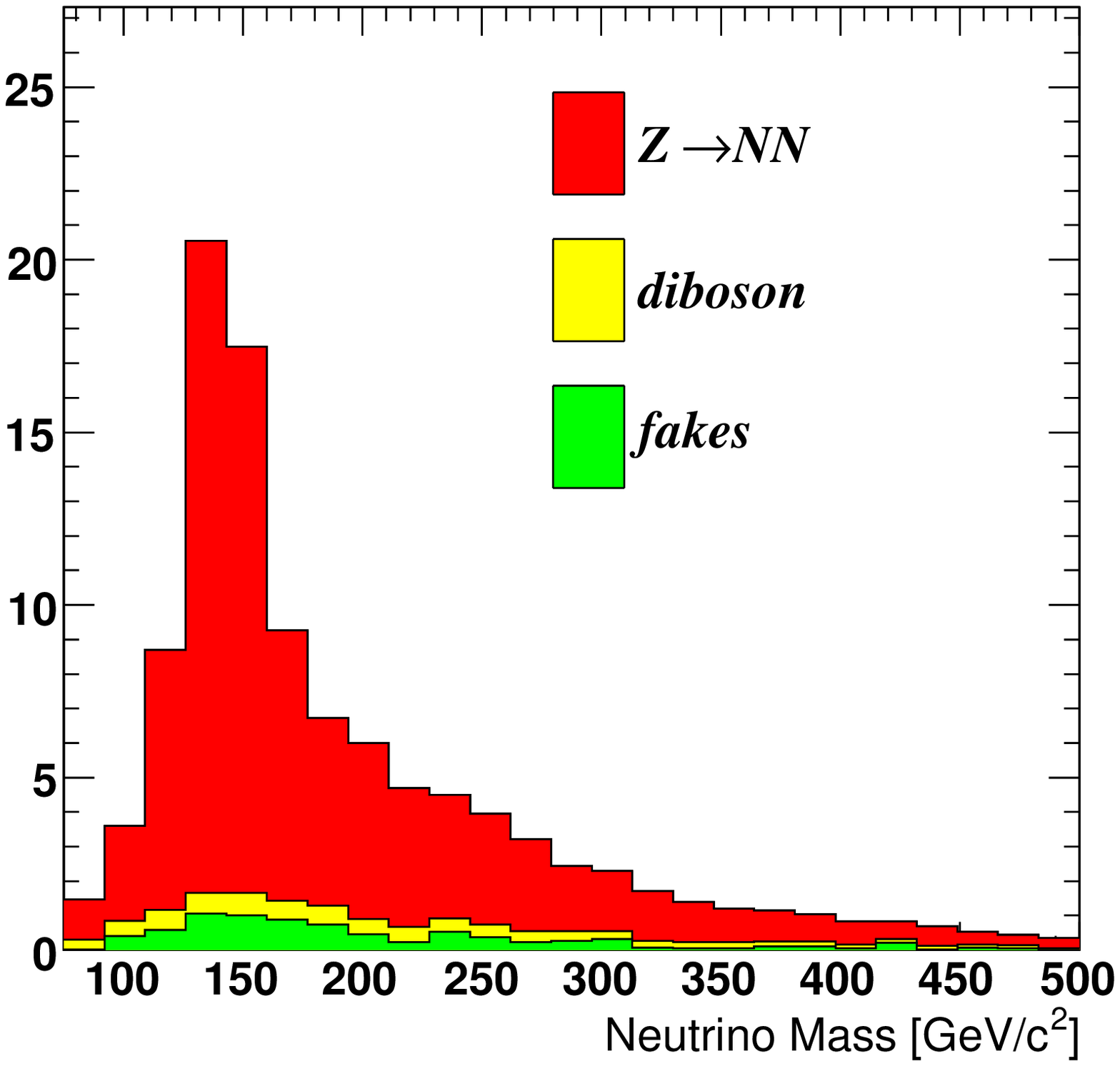}
\caption{Expected reconstructed neutrino mass for $N$ production
  with $m_{N}$ = 150 GeV/$c^2$, and
  backgrounds to the $\mu^{\pm}\mu^{\pm}jj$ signature in 5 fb$^{-1}$ of
  Tevatron data (top) or 10 TeV LHC data (bottom).}
  \vspace{-3ex}
\label{fig:sigback}
\end{figure}

Figure~\ref{fig:sigback} shows the reconstructed mass shape for $N$
pair production and for the backgrounds in the $\mu^{\pm}\mu^{\pm}jj$ case.

\subsection{Expected Limits and Discovery Potential}

\begin{figure}[h]
\includegraphics[width=0.45\linewidth]{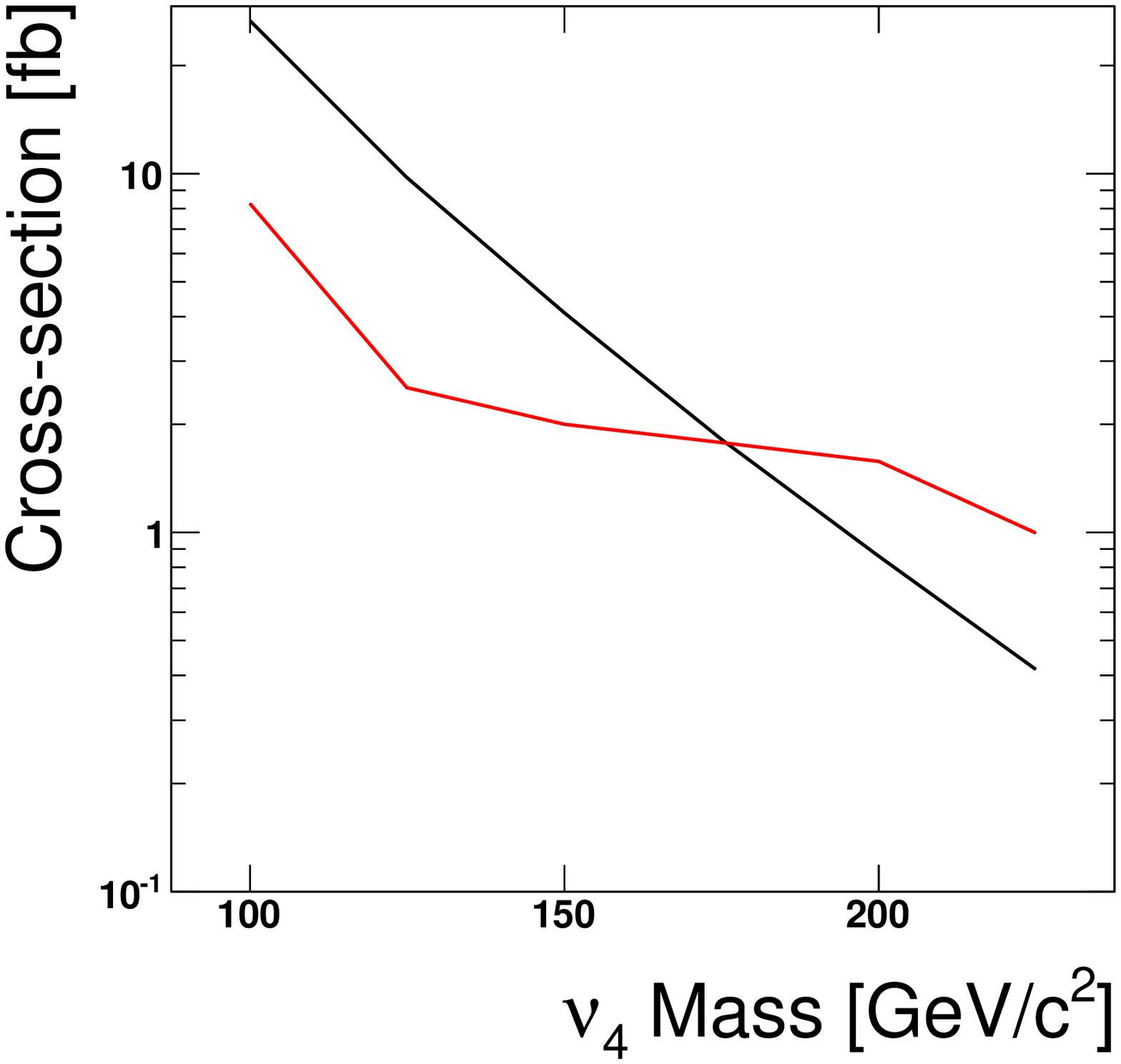}
\includegraphics[width=0.45\linewidth]{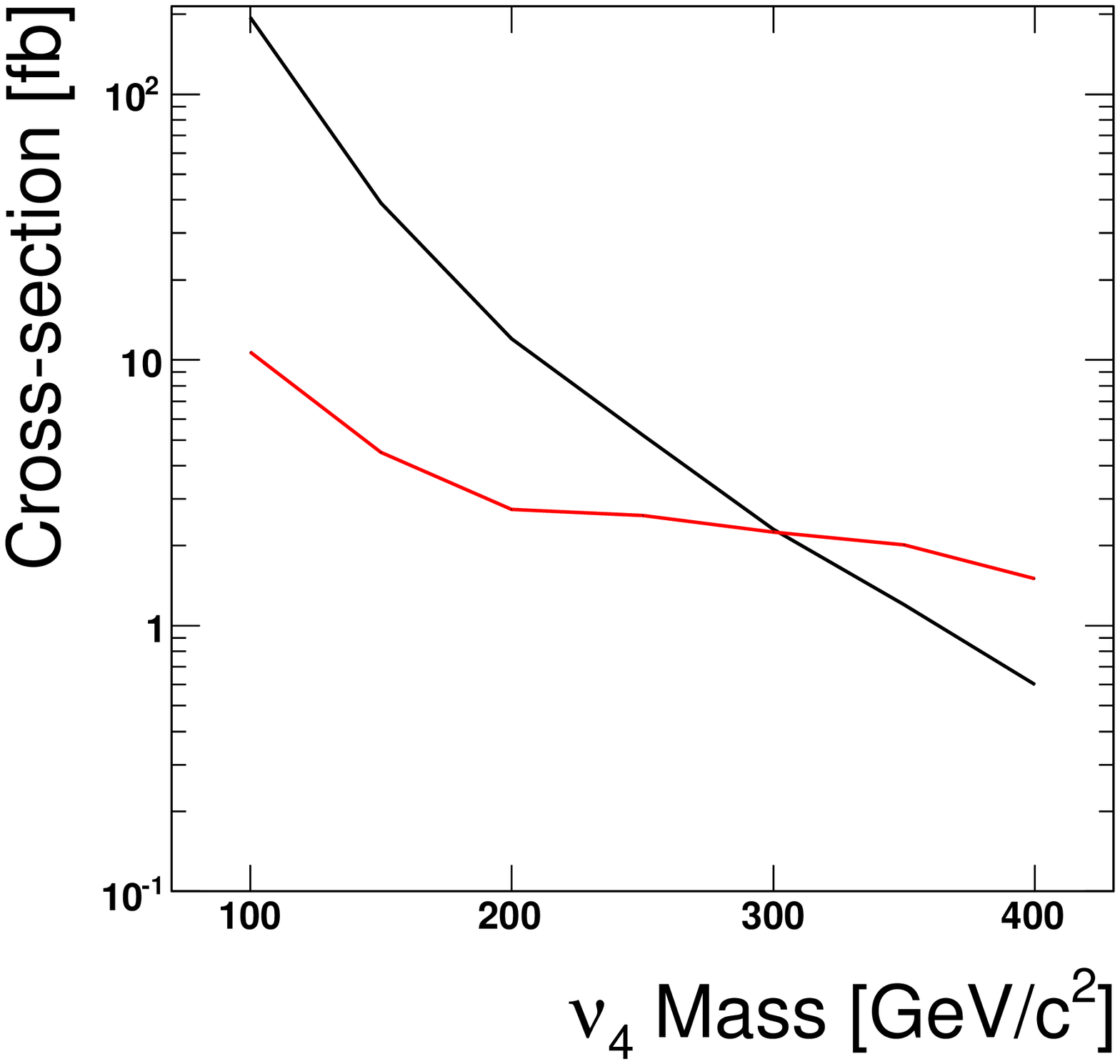}
\caption{Theoretical cross-section for $N$ production and decay to $\ell^{\pm}W\ell^{\pm}W$ and
  median expected 95\% C.L experimental cross-section upper limits in  5 fb$^{-1}$ of
  Tevatron data (left) or LHC data (right), assuming BR($N \rightarrow \mu W)=100$\%. }
  \vspace{-3ex}
\label{fig:sens}
\end{figure}

\begin{figure}[h]
\includegraphics[width=0.45\linewidth]{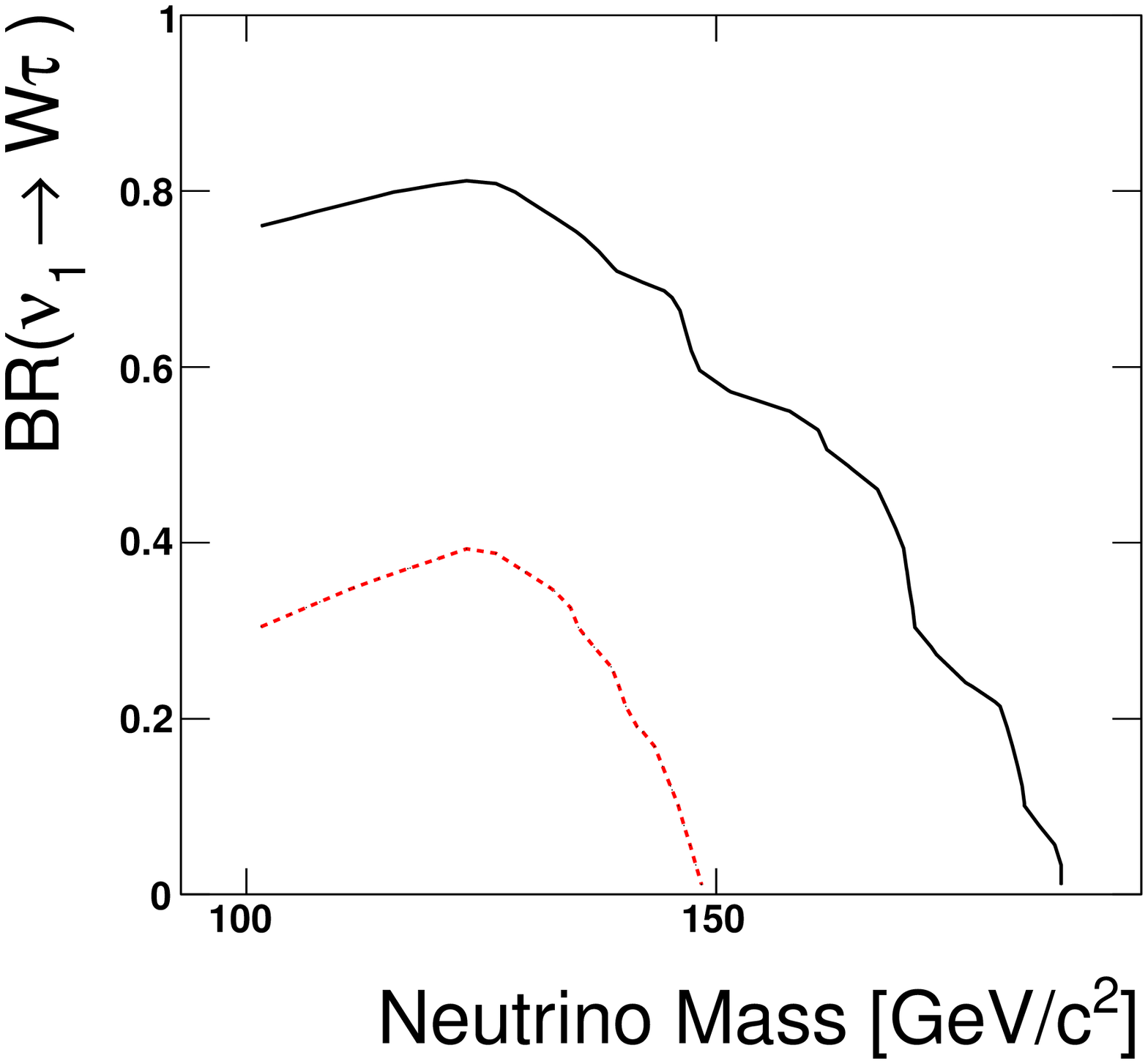}
\includegraphics[width=0.45\linewidth]{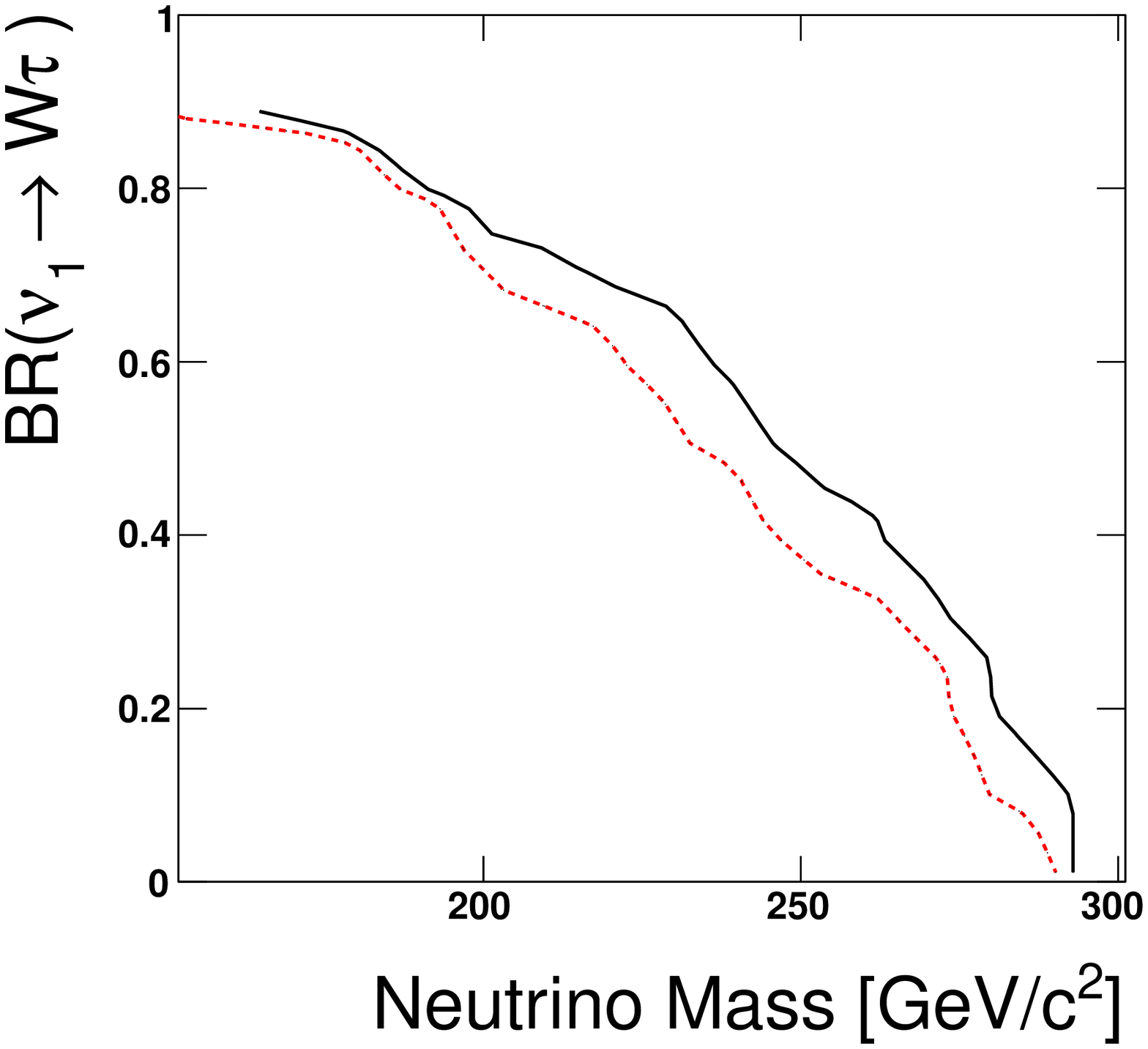}
\includegraphics[width=0.45\linewidth]{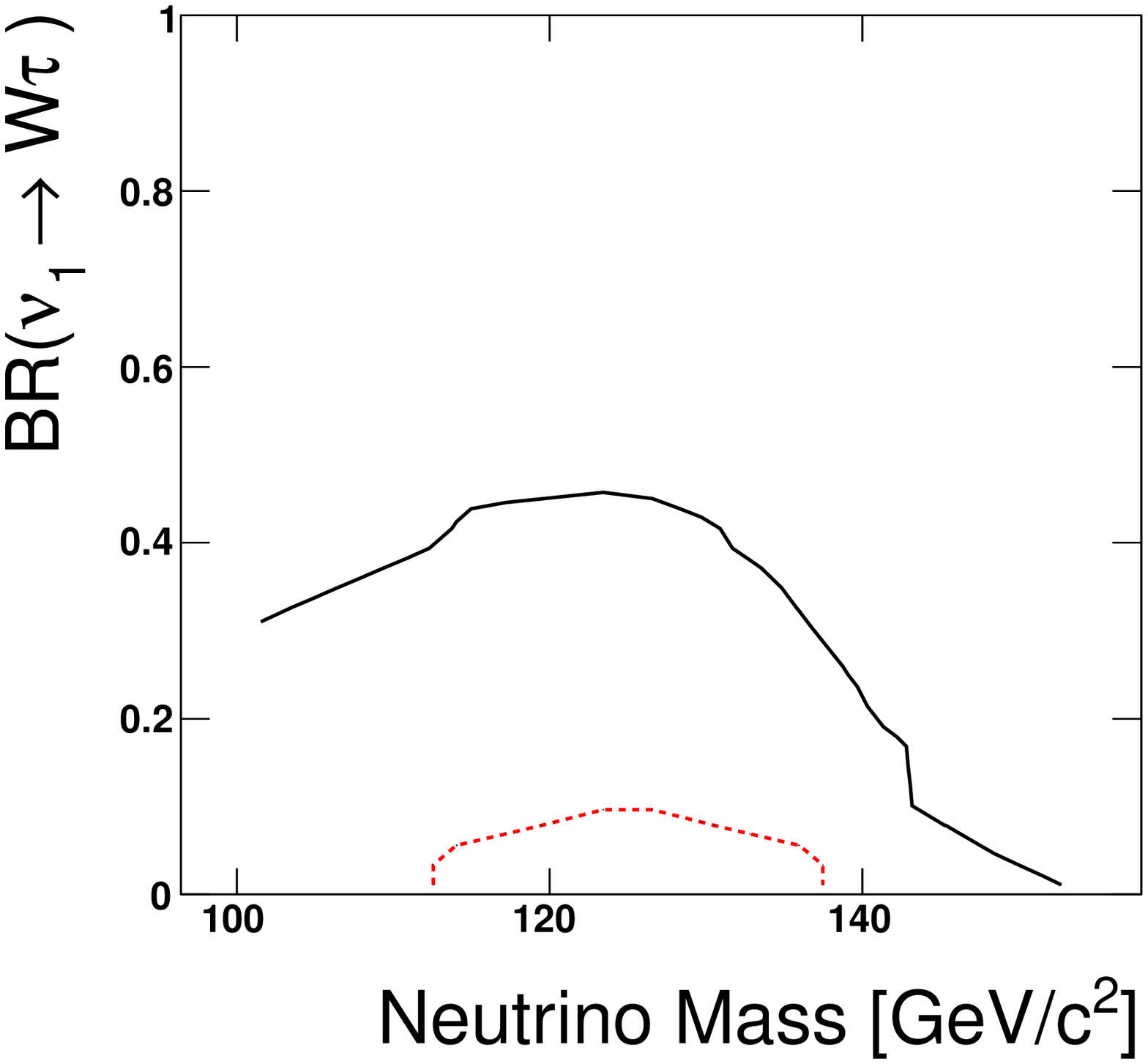}
\includegraphics[width=0.45\linewidth]{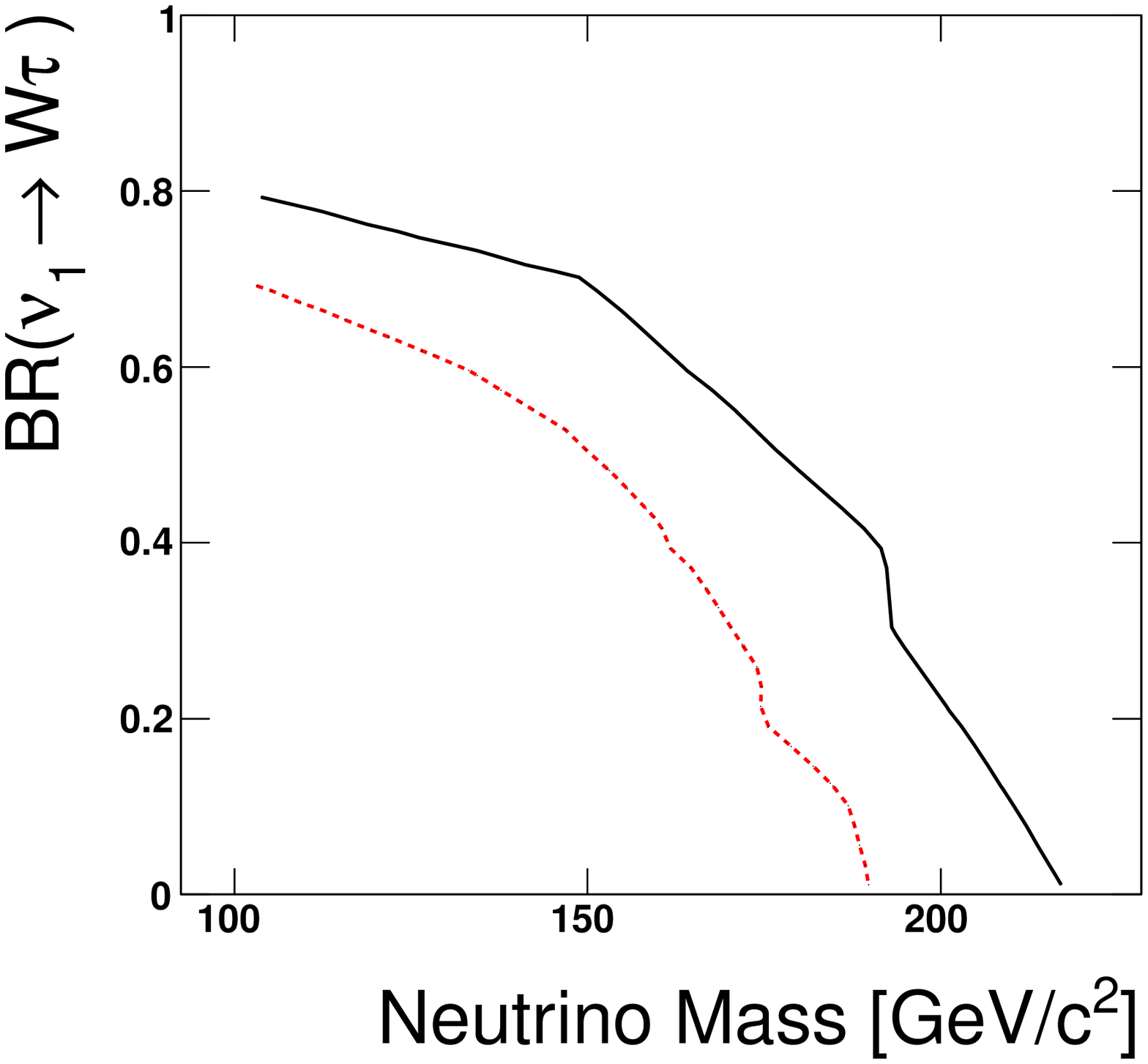}
\caption{ Median expected 95\% C.L experimental exclusion (top) or
  3$\sigma$ evidence (bottom) in 5 fb$^{-1}$ of
  Tevatron (left) LHC data (right) as a function of
  BR($N\rightarrow W\tau$).  Two decay cases are shown:
  $\mu^{\pm}\mu^{\pm}$  (black) or $\ell^{\pm}\ell^{\pm}$ (red), as defined in the
  text.}
  \vspace{-3ex}
\label{fig:mix}
\end{figure}

We perform a binned likelihood fit in the reconstructed $N$
mass, and use the unified ordering scheme~\cite{feldcous} to construct
frequentist intervals. If the $N$ does not exist and no excess is
seen, the median expected upper limits on the cross-section are given
in Table~\ref{tab:sens} and Fig. ~\ref{fig:sens}.  In 5 fb$^{-1}$,
with BR($N \rightarrow \mu W)=100$\%, a single Tevatron (LHC)
experiment could expect to set a 95\% lower limit of $m_{N}>175\ (300)$
GeV.  The limits as a function of   BR($N \rightarrow \tau W$) are
given in Fig.~\ref{fig:mix}.  If the $N$ does exist, a 3$\sigma$
excess would be observed in the regions shown in Fig.~\ref{fig:mix}.

\begin{table}
\caption{ Theoretical cross section, $\sigma_{Theory}$ at the Tevatron
  or LHC, including branching ratio to like-sign leptons; selection
  efficiency $\epsilon$ for the $\mu^{\pm}\mu^{\pm}jj$ channel; number of expected events in 5 fb$^{-1}$ of
  data; and median expected experimental cross section 95\% CL
 upper limits, $\sigma_{Limit}$, assuming BR($N\rightarrow
 W\mu$) = 100\%. }
\begin{ruledtabular}
\begin{tabular}{lrrrrrr}
\multicolumn{7}{c}{Tevatron} \\ \hline 
Mass [GeV/$c^2$] & 100 & 125 & 150 & 175 & 200 & 225 \\
$\sigma_{Theory}$[fb] & 26.7 & 9.8 & 4.1 & 1.8 & 0.9 & 0.4 \\
$\epsilon$ & 0.09 & 0.32 & 0.44 & 0.51 & 0.54 & 0.55 \\
Yield & 11.5 & 15.7 & 9.1 & 4.6 & 2.3 & 1.2 \\
$\sigma_{Limit}$[fb] & 8.3 & 2.5 & 2.0 & 1.8 & 1.6 & 1.0  \\
\end{tabular}
\begin{tabular}{lrrrrrrr}
\multicolumn{8}{c}{LHC, 10 TeV}\\ \hline
Mass [GeV/$c^2$] & 100 & 150 & 200 & 250 & 300 & 350 & 400 \\
$\sigma_{Theory}$[fb] & 195  & 39  & 12  & 5.2 & 2.3  & 1.2 & 0.6  \\
$\epsilon$ & 0.11 & 0.46 & 0.57 & 0.61 & 0.63 & 0.65 & 0.65 \\
Yield & 111.0 & 91.7 & 35.0 & 15.9 & 7.4 & 3.8 & 1.9 \\
$\sigma_{Limit}$[fb] & 10.7 & 4.5 & 2.7  & 2.6 & 2.3 & 2.0 & 1.5 \\
\end{tabular}
\end{ruledtabular}
\label{tab:sens}
\end{table}

\section{Conclusions}

The $s$-channel pair production of heavy Majorana  neutrinos via
a $Z$ boson ($Z\rightarrow NN \rightarrow W\ell W\ell$) is a powerful
discovery mode at hadron colliders.  
With 5 \ fb$^{-1}$ of data, the  Tevatron 
can significantly extend the limits on such neutrinos 
to 175 GeV/$c^2$, and a 3 $\sigma$ evidence is possible
if the mass is less than $ 150$\ GeV$/c^2$.  A dataset of the same size at
the LHC would have an 95\% C.L. exclusion reach of $300$ GeV$/c^2$ and
3$\sigma$ evidence potential for $m_N < 225$\ GeV$/c^2$.

\section{Acknowledgements}

We acknowledge discussions with M.~Bondioli, L.~Carpenter, K.~Slagle,
 T.~ Tait, and R.~Porter,. A.R. is supported in part by NSF Grant
No. PHY-0653656. D.W. is supported in part by the U.S. Dept. of Energy.

\end{document}